\documentclass[12pt,preprint]{aastex}
\usepackage{graphicx}
\usepackage{amssymb}
\newcommand{\ignore}[1]{}
\newcommand{\be}{\begin{equation}} \newcommand{\ee}{\end{equation}}
\newcommand{\ba}{\begin{eqnarray}} \newcommand{\ea}{\end{eqnarray}}
 \renewcommand{\bf}{\textbf}
\newcommand{\ra}{\rightarrow}

\def\slashb#1{\setbox0=\hbox{$#1$}#1\hskip-\wd0\dimen0=5pt\advance
        \dimen0 by-\ht0\advance\dimen0 by\dp0\lower0.5\dimen0\hbox
          to\wd0{\hss\sl/\/\hss}}

\begin{document}

\title{Evidence for Evolution or Bias \\ in Host Extinctions of Type 1a Supernovae at High Redshift   }
\author{ Pankaj Jain$^a$ and John P. Ralston$^b$\\
$^a${\it Physics Department, IIT, Kanpur - 208016, India}\\
$^b${\it Department of Physics \& Astronomy,}\\
{\it University of Kansas, Lawrence, KS - 66045, USA}\\
}


\begin{abstract}{Type 1a supernova magnitudes conventionally include an additive parameter called the extinction coefficient.  We find that the extinction coefficients of a popular ``gold'' set are well correlated with the deviation of magnitudes from Hubble diagrams.  If the effect is due to bias, extinctions have been overestimated, which makes supernovas appear more dim.   The statistical significance of the {\it extinction-acceleration correlation} has a random chance probability of less than one in a million.   The hypothesis that extinction coefficients should be corrected empirically 
provides greatly improved fits to both accelerating and non-accelerating models, with the independent feature of eliminating any significant correlation of residuals.  }
\end{abstract}



\section{Introduction}
Type 1a supernovas are candidates for standard astrophysical candles, from which the relation of redshift $z$ and distance can be estimated.  In a universe of constant expansion the ``Hubble plot'' made from magnitudes and redshifts should be a straight line.  Data is now available for a wide
range of redshifts up to 1.755 (Schmidt et al. 1998; Garnavich et al. 1998; 
Perlmutter et al. 1998; Riess et al. 1998; 
Perlmutter et al. 1999; Knop et al. 2003; Tonry et al. 2003; Barris et al 
2004).  The Hubble diagrams derived from supernovae have indicated an upward bending curve, interpreted as acceleration of the expansion rate, along with even more complicated features of ``jerk''.  It is important to explore other interpretations, including possible evolution of supernova or host galaxy 
characteristics with redshift.  Many
papers have explored non-cosmological explanations (Coil et al. 2000; Leibundgut 2001; Sullivan et al. 2003; 
Riess 2004). Meanwhile, the high redshift host galaxies have significantly different morphologies  
compared to those at low redshifts (Abraham \& van den Bergh 2001; 
Brinchmann et al. 1998; van den Bergh 2001). Dust and related extinction characteristics
may certainly depend on redshift (Totani \& Kobayashi 1999). 
Furthermore the abundance ratios of the progenitor stars may be
different at different redshifts (H\"{o}flich et al. 2000).  Several studies emphasize that evolution effects cannot be ruled out (Falco et al. 1999; Aguirre 
1999; Farrah et al. 2004;  Clements et al. 2004).  

In this paper we find evidence for evolution or bias in the extinction parameters used to pre-process the data.  If the effect is due to bias, extinctions have been overestimated, which makes supernovas appear more dim. Yet just the same phenomenon could occur from a real physical effect in which the actual host extinctions are correlated with the deviation of magnitudes from model fits. 

\subsection{Background} 

Traditional Hubble diagrams represent the relation of observed flux ${\cal F}$ to the luminosity of the source ${\cal L}$,  

\ba{ \cal F} ={{ \cal L} \over 4 \pi d_{L}^{2}}\ ,  \ea where $d_{L}$ is the so-called luminosity distance.  The distance modulus $\mu_p=m-M$, where $m$ and $M$ are the apparent and absolute magnitudes respectively, is 
\begin{equation}
\mu_p = 5\log d_L + 25 \ ,
\label{eq:distance_modulus}
\end{equation}
where the luminosity distance $d_L$ is in megaparsecs.  

The process of converting observed data into the supernova magnitudes reported actually contains an additive parameter, called the extinction coefficient $A$.  Extinction may depend on frequency, designated by $A_{B}$, $A_{R}$, etc.  The units of $A$ are magnitude.  In practice $A $ shifts the supernova magnitude $m_{0}$ deduced from light-curves to a reported magnitude (``extinction corrected  
magnitude'') $m= m_{0}-A $.  Our galaxy contributes extinction, as do the additional extinction effects associated with supernova host galaxies, which are more model dependent.  

\clearpage

\begin{figure}
\epsscale{.80}
\plotone{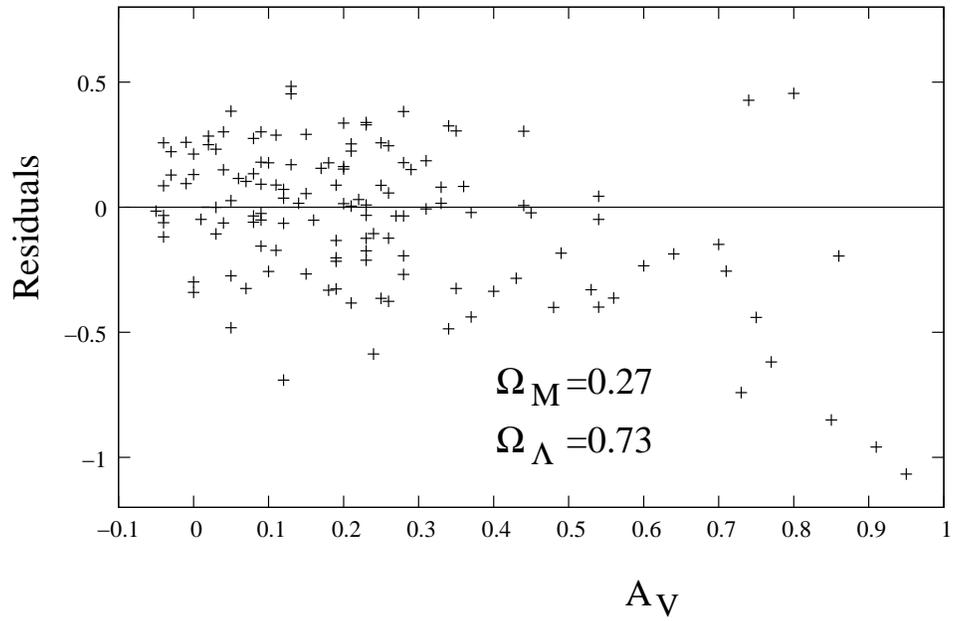}
\caption{The residuals as a function of the host extinction for the 
concordance model $\Omega_M=0.27$, $\Omega_\Lambda= 0.73$.
  }
\label{fig:residuals1}
\end{figure}
 
\clearpage

Riess {\it et al} (2004) discovered 16 Type Ia supernovas at high redshifts
and compiled a 157 source  ``gold'' data set held to be of the highest reliability. Extinctions are listed in Riess {\it et al} (2004) for all except 24
sources among this ``gold'' set.  

\section{Analysis} 

Riess {\it et al} focus on the differences of magnitudes $\Delta \mu$ relative to the traditional Hubble plot.  In Fig.
\ref{fig:residuals1} we show the residuals $\Delta \mu$ versus the 
extinction coefficients $A_V$, for all the sources for which extinctions 
are known. There is a clear correlation. The sense of correlation is that 
points with $\Delta \mu >0$, lying above the straight line Hubble plot, 
tend to have small or even negative extinction, and points lying below the 
straight line tend to have large extinction.  A precedent for examining 
correlations of residuals is given in Williams 
{\it et al.}, (2003). 

Residuals depend on the baseline model from which 
they are measured.  Fig.\ref{fig:residuals1} uses the 
FRW model and ``concordance'' 
parameters $\Omega_M=0.27$, with $\Omega_\Lambda= 0.73$ under 
the constraint $\Omega_k=0$.  This is one of the baselines cited by Riess {\it et al} (2004).  Here $\Omega_M$ is the matter density, $\Omega_\Lambda$ 
the vacuum energy density and $\Omega_k = 1-\Omega_M-\Omega_\Lambda$.  The 
class of FRW models predicts the luminosity distance as
\begin{equation}
d_L = {c(1+z)\over H_0 |\Omega_k|^{1/2}}{\rm sinn}\left\{|\Omega_k|^{1/2} 
\int_0^z dz
\left[(1+z)^2(1+\Omega_Mz) - z(2+z)\Omega_\Lambda\right]^{-1/2}\right\}
\label{eq:luminosity_distance}
\end{equation}
Here ${\rm sinn}$ denotes $\sinh$ for $\Omega_k>0$,  $\sin$ for $\Omega_k<0$
and is equal to unity for $\Omega_k=0$.  Parameters are fit by minimizing $\chi^{2}$, defined by  
\begin{equation}
\chi^2 = \sum_i  \,  {(\, \mu_p^i - \mu_0^i -\mu_{p0} \,)^{2} \over(\delta\mu_0^i)^{2}},
\end{equation} where $\mu_p^i$ and $\mu_0^i$ are the theoretical and observed
distance moduli respectively and $\delta\mu_0^i$ are the reported errors.   Our notation includes the intercept parameter $\mu_{p0}$ (not always explicit in the literature).  
The Hubble constant $H_0$ and fit parameters such as the zero point are not reported 
in Riess {\it et al} (2004), which states that they are irrelevant and arbitrarily 
set for the sample presented here.  We verify (Riess {\it et al}, 2004)
 $\chi^2=178$ for the concordance parameters cited above, along with the other $\chi^2$ values for several other studies, presented below.

\subsection{Quantification} 

We quantify the correlation of extinctions with residuals with the correlation
coefficient $R(\Delta \mu, A_{V})$, also simply $R$, defined by 
\ba  R(\Delta \mu, A_{V}) = {\sum_i   \,   (\, \Delta \mu_{i} -\bar  
\Delta \mu \, ) (\, A_{V, \,i}-\bar A_{V} \,) \over \sigma_{  \Delta \mu} 
\sigma_{A_V}  }, \ea 
where $\bar  \Delta \mu, \, \sigma_{\Delta \mu}$ are the means and standard 
deviation of the $\Delta \mu$ set, with corresponding meaning for 
$\sigma_{A_V}, \, \bar A_{V}$.  The correlation $R(\Delta \mu, A_{V}) =-0.439$
for the concordance parameters cited above, excluding the 24 sources for
which extinctions are not known.  The integrated probability 
(confidence level, $P$-value) to find correlations equal or larger in a 
random sample is $4.2\times 10^{-7}$.

To investigate whether the correlation of extinctions with residuals might be a model artifact, we decided to fit several other models cited by Riess {\it et al} (2004).  The results of these fits are shown in Table \ref{table_chi2}.  For example, under the best fit model with $\Omega_M=0.31, \Omega_\Lambda= 0.69 =1-\Omega_{M}$ then $R(\Delta \mu, A_{V}) =-0.434$ with probability $P= 5.6\times 10^{-7}$.

\clearpage

\begin{figure}
\plotone{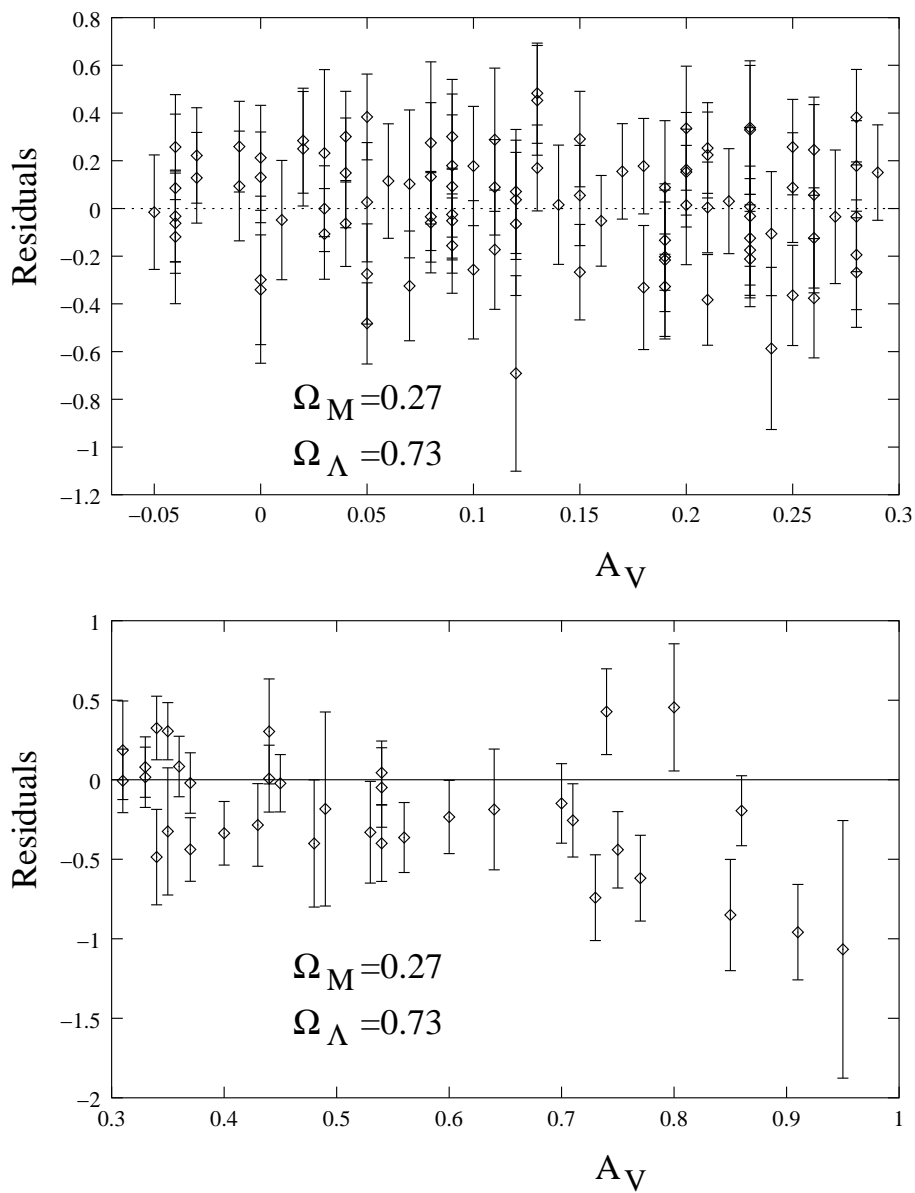}
\caption{  Residuals versus $A_{V}$, including reported uncertainties assigned to the residuals.  By inspection the region of $A_{V}  \gtrsim 0.3$ shows systematic correlation.   }
\label{fig:corErrorBars.eps}
\end{figure}

\clearpage

From Fig. \ref{fig:residuals1} we see that the correlation 
is strongest for large
values of $A_V$. For example, for the best fit parameters
$(\Omega_M=0.31, \Omega_\Lambda= 0.69)$ we find that excluding the
four sources with $A_V > 0.8$ the correlation coefficient goes down to 
$R(\Delta \mu, A_{V}) =-0.28$ with $P=1.5 \times 10^{-3}$.  Retaining the 
139 points with $A_{V}\le 0.5$ yields $R(\Delta \mu, A_{V}) = -0.18$.  
We do not have a particular reason to entertain these cuts except to make 
the correlation go away.  At the risk of complicating interpretation, one 
can try dividing the residuals by the data point's uncertainty. This is an uncertain trial because a fundamental issue is the {\it uncertainty in the extinction coefficients}, which is unavailable from the literature.  Fig. \ref{fig:corErrorBars.eps} shows the correlation with error bars assigned to the residuals.\footnote{We thank an anonymous referee for this suggestion.} The figure shows
that most of the data with $A_V>0.3$ lies below 0, indicating bias.
Division by uncertainty only reduces $ R(\Delta \mu/\sigma, \,  A_{V}) \ra -0.37$ for the gold set, an effect of having introduced noise.

We next examine whether the correlation seen in the residuals depends on 
redshift. We divide the data
as equally as possible in a large redshift sample ($z\ge 0.41$, 78 sources) 
and a low 
redshift sample ($z<0.41$, 79 sources).  (The cut $z \sim 0.46$ was identified 
by the Hubble team as a transition region.) For the  
low redshift sample we find $R(\Delta \mu, A_{V})  = -0.509$, 
$P = 1.2\times 10^{-5}$, compared to the high redshift sample yielding 
$R(\Delta \mu, A_{V})  = -0.378$, $P = 3.7\times 10^{-3}$. 
Although statistics have been diluted, it is clear that the two samples show
different behavior, with the correlation being much more significant in the low
redshift sample.  
\clearpage

\begin{table}
\begin{tabular}{|l|c|c|c|}
\hline
Model & $\chi^2$ & R & P\\
\hline
\hline
$\Omega_M=0.27$, $\Omega_\Lambda= 0.73$ & 178.2 & $-0.439$ & $4.2\times 10^{-7}$ \\ 
\hline
$\Omega_M=0.31$, $\Omega_\Lambda= 0.69$ 
& 177.1  & $-0.434$ & $5.6\times 10^{-7} $\\
(best fit with $\Omega_M+\Omega_\Lambda=1$) & & & \\
\hline
$\Omega_M=0.45$, $\Omega_\Lambda= 0.95$ (best fit)& 175.1 & $-0.403$ & $3.3\times 10^{-6}$\\
\hline
$\Omega_M=0.0$, $\Omega_\Lambda= 0.0$ (best fit with $\Omega_\Lambda= 0.0$)& 191.7 & $-0.392$ & $6.0\times 10^{-6}$\\
\hline
$\Omega_M=1$, $\Omega_\Lambda= 0$ & 324.8 & $-0.275$ & $1.49\times 10^{-3}$\\
\hline
\end{tabular}
\caption{ $\chi^2$ values, the correlation statistic $R(\Delta \mu, A_{V}) $ between residuals and extinction, and confidence level (P) to find 
$R(\Delta \mu, A_{V}) $ in a random sample. The different cosmological models
come from Riess {\it et al} (2004).}
\label{table_chi2}
\end{table}

\clearpage

Questions then branch along three lines: (1) The assignment of extinctions by present schemes may contain hidden bias.   (2) There may be a real physical effect at work, and (3) Systematic errors might be re-evaluated in order to ameliorate the significance of the correlation. 

\medskip 

 $\star$ 1:  A seldom discussed but established bias exists in the assignment of $A_{V}$ from the fits to light curves.  We find it highlighted by the Berkeley group (Perlmutter {\it et al} 1999, especially the Appendix).  The scheme used starts with a conditional probability $P(A|A_{dat})$, where $A_{dat} $ is the extinction from the best fit to the light curve data.  A prior probability $P_{0}(A_{dat} )$ is assumed, and from Bayes' Theorem the probability of $A$ after seeing the data is estimated. The value of $A$ is chosen to ``maximize the probability of $A$'' given the combined information from the prior and the data.  

The method introduces an extra dependence on the choice of priors.  For prior distributions centered at small host extinction, the work of Hatano (1998)
is cited, based on Monte Carlo estimates from host galaxies of random 
orientation.   Freedom is used to formulate a one-sided prior distribution 
with support limited to $A> 0$.   This make a bias in the combination of 
assuming $A>0$ for the priors (fluctuations could do otherwise) and the 
detailed way in which $A_{dat}$ is assigned.  This bias tends to cause 
the same signal as dimming or acceleration  (Perlmutter {\it et al} (1999)). 
As of 1999 the outcomes of this bias were stated to be less than 0.13 magnitude.  

Yet one would need an absolute standard to evaluate any bias reliably.  Subsequently the method itself has evolved (Riess et al. 2004), 
citing an iterative ``training procedure'' we have not found described 
in detail. A few points now have $A_{V}<0$.  

There is evidently a further bias in taking data from the peak of the proposed distribution. It is not the same thing as sampling the proposed distribution randomly.  Iteration of a procedure taking from the peak tends to drive a Bayesian update procedure towards a narrow distribution centered at the peak.  In some renditions this may cause systematic errors of fluctuations to evolve towards becoming underestimated.  

 $\star$ 2: It is possible that the extinction correlation is a signal of 
physical processes of evolution with redshift.  It is impossible to adequately summarize the literature discussing this possibility. Aguirre (1999) 
made a comparatively early study with a balanced conclusion that extinction models might cause some of the effects interpreted as acceleration. Drell, Loredo
\& Wasserman (2000) concentrate on this question, concluding that the methodology of using type 1a supernovas as standard candles cannot 
discriminate between evolution and acceleration. Farrah  et al. (2004) 
(see also Clements et al. 2004)
cite a history of work scaling optical frequency extinction with the sub-millimeter wavelength observations (Hildebrand 1983; Casey 1991; Bianchi 1999). 
They report extinction for 17 galaxies with $z=0.5$ with sub-millimeter 
wavelengths.  While stating consistency with local extinctions at the $1.3 
\sigma$ level, they add ``It does however highlight the need for caution 
in general in using supernovae as probes of the expanding Universe, as 
our derived mean extinction, $A_{V}= 0.5 \pm 0.17$, implies a rise that is
{\it at face value} comparable to the dimming ascribed to dark energy. 
Therefore, our result emphasizes the need to accurately monitor the extinction
 towards distant supernovae if they are to be used in measuring the 
cosmological parameters.''  The trend of Farrah's observation is same as 
the correlation seen in the supernova data, and remarkably, the corrections
we obtain empirically in various fits (below) almost all amount to 0.5
magnitude or less. The fact that low redshift objects show higher correlation
 implies that there is a higher tendency to overestimate
extinctions of these sources in comparison to the sources at higher redshifts.  
Since the estimated extinctions show no correlation with redshift, this
suggests that the true low redshift extinctions, on the average, may be
smaller in comparison to the extinctions of high redshift sources.  
Nevertheless the question of 
evolution of the sources remains open and will not be resolved here.  

  $\star$ 3:  Perhaps the means of assigning extinction coefficients are reasonable on average, but statistical fluctuations have given a false signal. Then the error bars on the extinction coefficients come to be re-examined.  Inasmuch as this is coupled to the entire chain of data reduction, it is beyond the scope of this paper.

\subsection{Empirically Corrected Extinctions}

Without engaging in physical hypotheses of extinction, it is reasonable to test whether a different extinction model can give a satisfactory fit to the data.    
We studied a corrected value $A_V(\delta) $ depending on the  parameter 
$\delta$
by the simple rule \begin{equation}
A_V(\delta)= (1+\delta)A_V\ . 
\label{eq:corrected_AV1}
\end{equation}
We then determine
$\delta$ by the best fit to the cosmological model.  The best fit $\delta$-values and the corresponding $\chi^2$ values for
different models are given in Table \ref{table_chi2_2}.  Parameter $\delta$ produces a huge effect of more than 23 units of $\chi^{2}$. 

There are many ways to compare the new and old fits. As a rule, 
the model with $\chi^{2}$ per degree of freedom ($dof$, the number of data
points minus the number of parameters) closest to unity is favored.  Since the new fits decrease $\chi^{2}$ by 20-some units with one additional parameter, 
the significance of revising the extinction values is unlikely to be fortuitous.
For example the model with $\Omega_M=0.27$ and $\Omega_\Lambda=0.73$
gives $\chi^{2}/dof = 1.14$ and $1.01$ without correction ($\delta=0$) and with correction ($\delta=-0.42$). As a broad rule in comparing data sets, the difference $\Delta \chi^{2}$ should be distributed by $\chi^{2}_{\nu}$, where $\nu=1$ is the number of parameters added.  The naive $p$-value or confidence level to find $\Delta \chi^{2} =23$ in $\chi^{2}_{1}$ is 1.6 $\times 10^{-6}$.  Thus introducing $\delta$ would be well-justified simply to improve the poor fit of $\chi^{2}\sim 178$ without ever seeing the extinction correlation with residuals.  Values of $\delta$ for all models are found to be negative, suggesting that the host extinction values given in Riess {\it et al} (2004) are overestimates.  

\begin{figure} 
\plotone{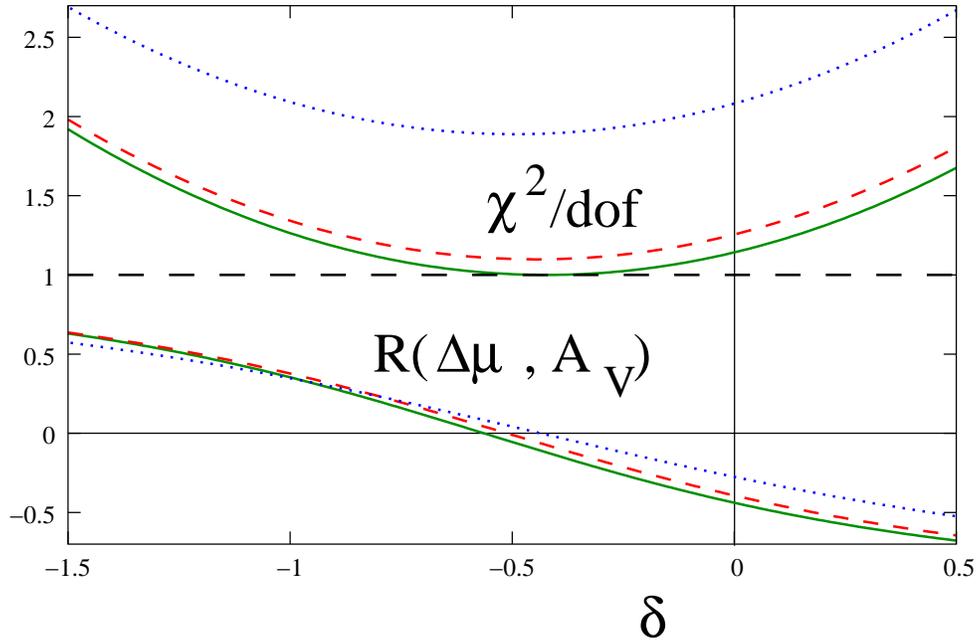}
\caption{\small  Best fit values of 
$\chi^{2}/dof$ versus parameter $\delta$ (upper curves), $\Omega_{M}=0.27$ 
(solid, green online), \, 0.5 (dashed, red online), \, 1 
(dotted, blue online) along with extinction correlation 
$R(\Delta \mu, A_{V}) $ (lower curves). It is significant that $R(\Delta \mu, A_{V}) $ crosses zero just in the vicinity of the best fit $\delta$ parameter.  } \label{fig:Chi2AndCorrel} \end{figure}

It is interesting and significant that the new residuals, computed relative to
the revised fits, show negligible correlation with host extinction.  This is seen in Fig. \ref{fig:Chi2AndCorrel}, which shows the $R$ values on the same plot as $\chi^{2}/dof$.  The fact that $R$ vanishes when $\delta$ meets the best-fit value is significant.  It is far from trivial, as $R$ concerns an independent set of numbers, the $A_{V}$ values, not directly used in calculating $\chi^{2}$. 

\begin{figure}


\plotone{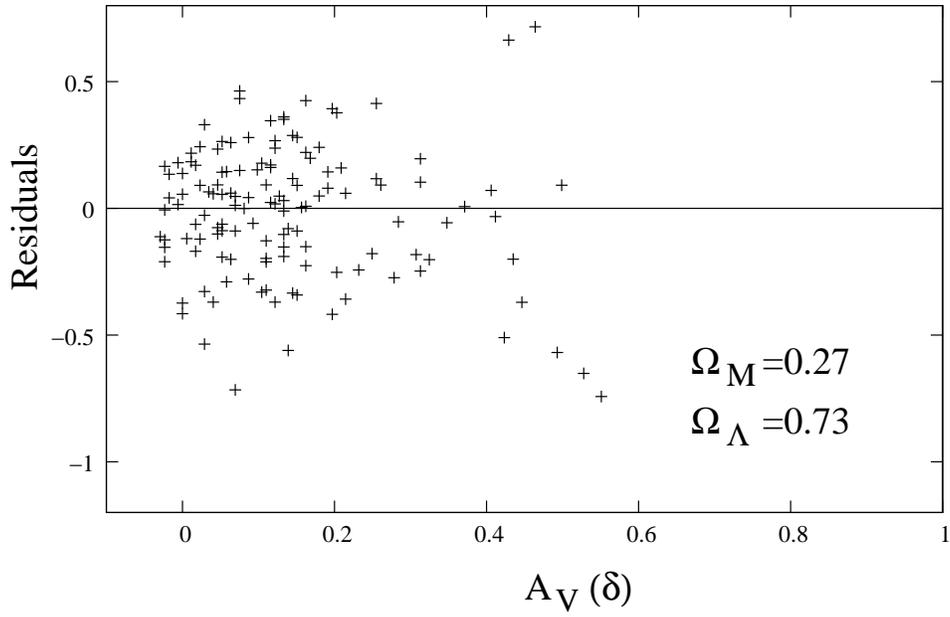} 
\caption{Magnitude residuals $\Delta \mu$ versus host extinction after 
including a correction term $A_V(\delta)= (1+\delta)A_V$ (Eq. \ref{eq:corrected_AV1}) in the distance modulus.  Parameters $\Omega_M=0.27$, $\Omega_\Lambda= 0.73=1-\Omega_M$, $\delta=-0.42$, as in Table \ref{table_chi2_2}.   }
  \label{fig:residuals2}
  \end{figure}

Figure \ref{fig:residuals2} shows the residuals versus corrected host extinction after 
including the correction term.  The reduction in correlation $R$ comes with an increased scatter in $A_{V}(\delta)$ at large $A_{V}(\delta)$, which is not unexpected.   

\clearpage

\begin{table}
\begin{tabular}{|l|c|c|c|c|}
\hline
Model&$\delta$ &$\chi^2$& R & P\\
\hline
\hline
$\Omega_M=0.27$, $\Omega_\Lambda= 0.73$ & $-0.42$  & 156.0 &$-0.12$ & 0.15\\
\hline
$\Omega_M=0.32$, $\Omega_\Lambda= 0.68$ 
& $-0.43$ & 154.5 & $-0.10$ &0.23 \\
(best fit with $\Omega_M+\Omega_\Lambda=1$) & & & &\\
\hline
$\Omega_M=0.35$, $\Omega_\Lambda= 0.75$ (best fit)&$-0.42$ &154.4 &$-0.11$
 &0.22 \\
\hline
$\Omega_M=0.0$, $\Omega_\Lambda= 0.0$ (best fit with $\Omega_\Lambda= 0.0$) 
& $-0.49$ & 162.5 & $-0.11$ & 0.20\\
\hline
$\Omega_M=1$, $\Omega_\Lambda= 0$&$-0.49$ & 294.6 & 0.04&0.68 \\
\hline
\end{tabular}
\caption{The $\chi^2$ values for some cosmological models including a
correction term in the distance modulus $A_{V}(\delta)=(1+\delta) A_V$ (Eq. \ref{eq:corrected_AV1}) due to possible bias in the 
host extinction.  Also shown are the correlation statistic $R(\Delta \mu, A_{V}) $ between residuals and extinction, and confidence level (P) to find $R$ in a random sample. }
\label{table_chi2_2}
\end{table}

\clearpage

\clearpage

\clearpage

It is also interesting to ask whether host extinction might have some
dependence on the luminosity distance $d_{L}$.  It is hard to imagine no evolution at all, and we explored a linear ansatz.   The linear model is \begin{equation}
 A_V(\delta, \, d_{L})=(1+\delta) A_V+  \delta_{1} d_{L}\ . \label{Correction2}
\end{equation}  We add that when a model of evolution is introduced, the cosmological interpretation might be disturbed, so that the outcomes must be taken in context.  More cannot be anticipated because the fits themselves will choose $\delta_{1}$.  Fit parameters and $\chi^2$ values are given in Table \ref{table_chi2_3}. 

\clearpage  
      
\clearpage

\subsubsection{Is Acceleration Supported?}  

\begin{figure}
\plotone{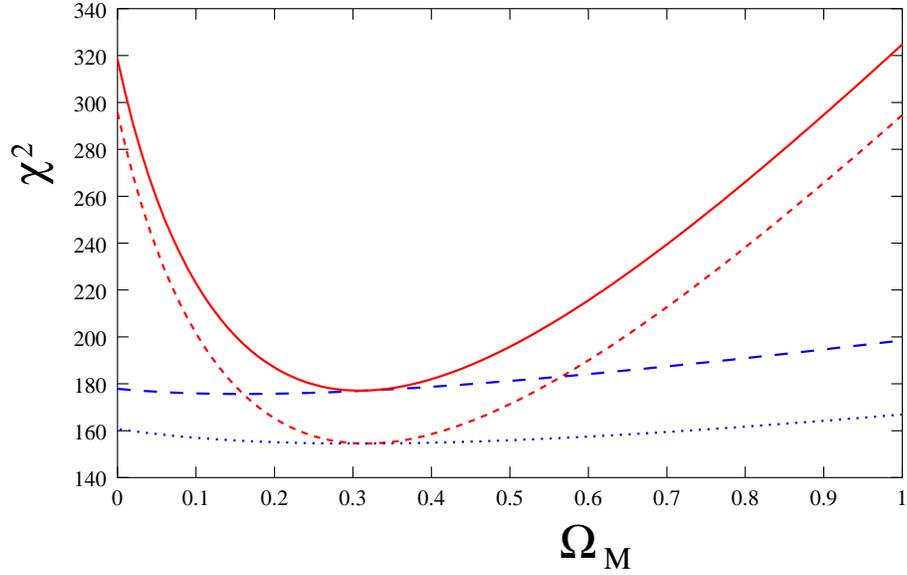}
\caption{The $\chi^{2}$ values versus $\Omega_{M}$ parameter in various models:
with uncorrected extinctions ($\delta =0$, solid, red online), best fit 
$A_V(\delta) = (1+\delta)A_{V}$ (short dashed, red online), best fit 
$A_V(d_{L}) = A_V + \delta_{1}  d_{L}  $ (long dashed, blue online), 
and best fit $A_V(\delta, \, d_{L})=(1+\delta) A_V+  \delta_{1} 
 d_{L} $ (dotted curve, blue online). The models with 
$\delta_{1} \neq 0$ are less sensitive to the value of $\Omega_{M}$. 
}
\label{fig:ChiSquared3Models}
\end{figure}

Accelerating models show no need for the $\delta_{1}$ term.  Assuming acceleration, the fits (Table \ref{table_chi2_3}) show that reducing extinction values by about 40\% explains the data better, and removes an alarming correlation.  On the other hand the matter-dominated model $(\Omega_M=1$, $\Omega_\Lambda= 0)$ shows interesting sensitivity to $\delta_{1}$.  In Fig.\ref{fig:ChiSquared3Models} we compare the sensitivity of different fits to parameter $\Omega_{M}$.  With $\delta_{1}=0$ constrained, the effects of $\delta$ are rather orthogonal to those of $ \Omega_{M}$, so that the region $\Omega_{M} \sim 0.3$ is favored whether or not there is a significant correlation $R$.  
Yet varying $\delta_{1}$ greatly broadens acceptable values of $\Omega_{M}$, while maintaining the $R \ra 0$ effect of $\delta$. The significance depends on one's hypothesis: if one chooses $\Omega_{M}=1$ {\it a-priori}, parameter $\delta_{1}$ is traded for parameter $\Omega_{M}$.  The overall probability of either hypothesis is only in part determined by the $p$-value of the data given the distribution: the rest depends on one's prior beliefs in evolution, which we will not pursue.  It is fair to say that the revised fits give more leeway to matter-dominated models on statistical grounds.   

In all cases fits are driven to $A_{V} \ra A_{V}(\delta \sim  -0.4) \sim 0.6 A_{V}  $, either simply to improve $\chi^{2}/dof$, or to remove the correlation with residuals.   

To conclude, analysis using reported extinction coefficients is well known to 
produce good fits to acceleration of the expansion rate. However the extinctions show correlation
with residuals with random chance probability using two 
independent tests, the extinction correlation and $\chi^{2}$ values,  
both below the level of $10^{-6}$. The hypothesis that extinction 
coefficients should be corrected empirically 
provides substantially improved fits to the data, while also eliminating significant correlation of residuals.  A model of linear evolution yields interesting
effects of high statistical significance correlated with redshift.  The studies indicate either 
bias in host extinction assignments or evolution of the source galaxies.  The significance of acceleration itself cannot be resolved on the basis of 
these studies, but might be revised, depending on one's priors.   We suggest that observers report uncertainties in their assignment of extinction parameters, both in the future and for the existing data sets.

\clearpage

\begin{table}
\begin{tabular}{|l|c|c|c|c|c|}
\hline
Model&$\delta$ & $\delta_1$& $\chi^2$ & R & P\\
\hline
\hline
$\Omega_M=0.27$, $\Omega_\Lambda= 0.73$ & $-0.42$ &$-0.037$ & 154.6 &$-0.11$
&0.18 \\
\hline
$\Omega_M=0.31$, $\Omega_\Lambda= 0.69$ 
& $-0.42$ & $-0.007$ & 154.5 & $-0.11$ & 0.20 \\
(best fit with  $\Omega_M+\Omega_\Lambda=1$) & & & & & \\
\hline
$\Omega_M=0.68$, $\Omega_\Lambda= 0.82$ (best fit) & $-0.42$ & 0.16 & 154.0
& $-0.079$ & 0.36 \\
\hline
$\Omega_M=0.0$, $\Omega_\Lambda= 0.0$ 
& $-0.48$ & 0.005& 162.4 & $-0.11$ & 0.19\\
(best fit with $\Omega_\Lambda= 0.0$)& & & & & \\ 
\hline
$\Omega_M=1$, $\Omega_\Lambda= 0$&$-0.51$& 0.47  & 166.9 & $-0.03$ & 0.73 \\
\hline
\end{tabular}
\caption{The $\chi^2$ values for some cosmological models, including a
correction term $A_V(\delta, \, d_{L})=(1+\delta) A_V+  \delta_{1} d_{L}$ (Eq. \ref{Correction2}) in the distance modulus due to possible bias. }
\label{table_chi2_3}
\end{table}

\clearpage

\clearpage

Acknowledgments:  Research supported in part under DOE Grant Number 
DE-FG02-04ER14308.  This work was completed when PJ was visiting 
the National Center for Radio Astrophysics, Pune. He thanks Prof. V.   
Kulkarni for kind hospitality. JP thanks Hume Feldman and Ruth Daly for discussions.

\bigskip
\noindent
\bf {References}
\bigskip

\begin{itemize}

\item[] Abraham, R. G., \& van den Bergh, S. 2001, Science, 293, 1273

\item[]  Aguirre, A. 1999, ApJ, 525, 583  

\item[] Barris, B., et al. 2004, ApJ, 602, 571

\item[] Bianchi, S., Davies, J. L., and Alston, P.B. 1999 A\&A, 344, L1. 

\item[] Brinchmann, J., et al. 1998, ApJ, 499, 112

\item[] Casey, S.C. 1991, ApJ, 371, 183 

\item[] Clements, D. L., Farrah, D., Fox, M., Rowan-Robinson, M., 
Afonso, J. 2004, New Astron. Rev. 48, 629 

\item[] Coil, A. L., et al. 2000, ApJ, 544, L111

\item[] Drell, P. S.; Loredo, T. J.; Wasserman, I. 2000,
ApJ, 530, 593.

\item[] Falco, E., et al. 1999, ApJ, 523, 617

\item[] Farrah, D., Fox, M., Rowan-Robinson, M., Clements, D., 
Afonso, J.  2004, ApJ, 603, 489  

\item[] Garnavich, P. M., et al. 1998, ApJ, 493, L53

\item[] Hatano, K., Branch, D. and  Deaton, J. 1998, ApJ, 502, 177

\item[] Hildebrand, R. H. 1983, QJRAS, 24, 267 

\item[] H\"{o}flich, P., Nomoto, K., Umeda, H., \& Wheeler, J. C. 2000, ApJ, 528, 
590

\item[] Knop, R., et al. 2003, ApJ, 598, 102

\item[] Leibundgut, B. 2001, ARA\&A, 39, 67

\item[] Perlmutter, S., et al. 1998,   [Supernova Cosmology 
Project Collaboration], Nature, 391, 51 

\item[] Perlmutter, S., et al. 1999  [Supernova Cosmology 
Project Collaboration], ApJ,   517, 565. 

\item[] Riess, A. G., et al 1998, AJ, 116, 1009 

\item[] Riess, A. G.,  et.al. 2004, ApJ, 607, 665

\item[] Riess, A. G. 2004, PASP, 112, 1284 

\item[] Schmidt, B. P., et al. 1998, ApJ, 507, 46

\item[] Sullivan, M., et al. 2003, MNRAS, 340, 1057

\item[] Totani, T., \& Kobayashi, C. 1999, ApJ, 526, L65 

\item[] Tonry, J. T., et al 2003, ApJ, 594, 1

\item[] van den Bergh, S. 2001, AJ, 122, 621

\item[] Williams, B. F. et.al. 2003 AJ, 126, 2608.

\end{itemize}
\end{document}